\documentclass[10pt,conference]{IEEEtran}
\IEEEoverridecommandlockouts
\usepackage{cite}
\usepackage{amsmath,amssymb,amsfonts}
\usepackage{algorithm}
\usepackage[noend]{algpseudocode}
\usepackage{graphicx}
\usepackage{textcomp}
\usepackage{xcolor}
\usepackage{multirow}
\usepackage{subcaption}
\usepackage{xfrac}
\usepackage{outlines}
\usepackage{siunitx}
\usepackage{braket}
\usepackage{mathtools}
\usepackage[normalem]{ulem}

\makeatletter
\newcommand{\linebreakand}{%
  \end{@IEEEauthorhalign}
  \hfill\mbox{}\par
  \mbox{}\hfill\begin{@IEEEauthorhalign}
}
\makeatother

\usepackage{changes}

\definechangesauthor[name="Alex", color=red]{ap}
\definechangesauthor[name="Hanjing", color=blue]{hx}
\definechangesauthor[name="Ilya", color=red]{is}

\def\BibTeX{{\rm B\kern-.05em{\sc i\kern-.025em b}\kern-.08em
    T\kern-.1667em\lower.7ex\hbox{E}\kern-.125emX}}
\begin{document}

\title{QAOA Parameter Transferability for Maximum Independent Set using Graph Attention Networks
\\}

\author{
\IEEEauthorblockN{Hanjing Xu}
\IEEEauthorblockA{\textit{Department of Computer Science} \\
\textit{Purdue University}\\
West Lafayette, Indiana USA \\
xu675@purdue.edu}\\

\IEEEauthorblockN{Alex Pothen}
\IEEEauthorblockA{\textit{Department of Computer Science} \\
\textit{Purdue University}\\
West Lafayette, Indiana USA \\
apothen@purdue.edu}

\and

\IEEEauthorblockN{Xiaoyuan Liu}
\IEEEauthorblockA{\textit{Fujitsu Research of America} \\
Santa Clara, California USA \\
xliu@fujitsu.com}\\

\IEEEauthorblockN{Ilya Safro}
\IEEEauthorblockA{\textit{Department of Computer and Information Sciences} \\
\textit{University of Delaware}\\
Newark, Delaware USA \\
isafro@udel.edu}

}
\maketitle

\begin{abstract}
The quantum approximate optimization algorithm (QAOA) is one of the promising variational approaches of quantum computing to solve combinatorial optimization problems. In QAOA, variational parameters need to be optimized by solving a series of nonlinear, nonconvex optimization programs. In this work, we propose a QAOA parameter transfer scheme using Graph Attention Networks (GAT) to solve Maximum Independent Set (MIS) problems. We prepare optimized parameters for graphs of 12 and 14 vertices and use GATs to transfer their parameters to larger graphs. Additionally, we design a hybrid distributed resource-aware algorithm for MIS (HyDRA-MIS), which decomposes large problems into smaller ones that can fit onto noisy intermediate-scale quantum (NISQ) computers. We integrate our GAT-based parameter transfer approach to HyDRA-MIS and demonstrate competitive results compared to KaMIS, a state-of-the-art classical MIS solver, on graphs with several thousands vertices.\\
{\bf Reproducibility: } Our source code and data are available at [link will be available upon acceptance].
\end{abstract}

\section{Introduction}\label{sec:intro}

Quantum computing is rapidly advancing as a powerful technology with substantial potential across a range of fields, including finance~\cite{herman2023quantum}, chemical simulations~\cite{cao2019quantum}, combinatorial optimization~\cite{shaydulin2019hybrid}, and machine learning~\cite{biamonte2017quantum}, among others. Presently, we are in the \emph{noisy intermediate-scale quantum} (NISQ) era, marked by quantum devices with tens to thousands of noisy qubits, limited coherence times, and restricted qubit connectivity.

The Quantum Approximate Optimization Algorithm (QAOA)~\cite{farhiQuantumApproximateOptimization2014} is one of the leading candidate algorithms for demonstrating quantum advantages in solving combinatorial optimization problems for NISQ and future fault-tolerant devices. It is a hybrid algorithm where parameters in the circuits are updated via classical feedback loops. The performance of QAOA depends on the choices of its parameters, which may be time-consuming to compute. Several previous results have shown concentrations in the parameter space for specific problems~\cite{wurtzFixedAngleConjecture2021,streifTrainingQuantumApproximate2019}, enabling parameter transfer from a set of pre-computed problem instances to new instances in unweighted MaxCut problems~\cite{lotshawEmpiricalPerformanceBounds2021,galdaTransferabilityOptimalQAOA2021,wurtzFixedAngleConjecture2021,brandaoFixedControlParameters2018} and weighted MaxCut problems~\cite{shaydulinParameterTransferQuantum2023}. However, many problems of practical interests have constraints, which may introduce new challenges for existing transfer schemes.

We introduce a parameter transfer scheme using Graph Attention Networks (GAT)~\cite{velickovicGraphAttentionNetworks2018, brodyHowAttentiveAre2022} for solving Maximum Independent Set (MIS) problems using QAOA. Additionally, to solve MIS problems on graphs with several thousands of vertices which are beyond the capabilities of NISQ computers, we partition the graph into contiguous components of small enough sizes, and solve each subproblem with QAOA in a distributed manner. The solution is then iteratively refined using different partitions for several rounds.
Given the current uncertainty surrounding which quantum hardware technology will ultimately prevail and scale effectively, it is essential to invest in research on \textit{distributed} hybrid quantum-classical algorithms with the main driving routine on classical high-performance computing machine reinforced with  multiple smaller-scale quantum devices for solving parts of the global optimization problem. Such algorithms offer the potential to integrate multiple types of quantum architectures, enabling collaborative computation across heterogeneous platforms. By minimizing the need for inter-device communication, these approaches can mitigate hardware-specific limitations and improve the overall scalability  of quantum computing solutions for large-scale problems.

In this work, we introduce the GAT-based parameter transfer scheme to MIS problems by learning the hidden connections between graph structures and their optimized QAOA parameters. Given a graph $G=(V,E)$, an \textit{independent set} in $G$ is a subset of its vertices where no two vertices are connected. The MIS problem is to find an independent set of maximum cardinality, and it is proven to be NP-hard~\cite{10.5555/574848} to solve exactly or approximately. MIS problems have applications in various areas including covering problems~\cite{brotcorneFastHeuristicsLarge2002}, macromolecular docking~\cite{gardinerCliquedetectionAlgorithmsMatching1997} and genome mapping analysis~\cite{harleyUniformIntegrationGenome2001}. In order to solve larger instances beyond hardware or simulation capabilities, we integrate the parameter transfer scheme into our distributed algorithm, where graphs are reduced to smaller instances whose parameters are transferred from a pool of pre-optimized QAOA circuits. We will compare independent set cardinalities computed by our algorithm with those from KaMIS~\cite{DBLP:journals/heuristics/LammSSSW17,DBLP:conf/alenex/Lamm0SWZ19}, a state-of-the-art classical solver and another hybrid separator-based divide-and-conquer (DC) algorithm~\cite{xuDivideConquerbasedQuantum2024}.

\noindent {\bf Our contributions:}
\begin{outline}
    \1 We design a framework that uses GATs to transfer QAOA parameters between MIS instances. Previous studies on QAOA parameter transfer focus mainly on MaxCut problems, whereas many problems in practice are constrained. To the best of our knowledge, this is the first attempt to use GATs, or Graph Neural Networks in general, for QAOA parameter transfer.
    \1 We propose a hybrid distributed resource-aware MIS algorithm, or HyDRA-MIS, that can solve MIS instances beyond the capabilities of existing quantum hardware. The algorithm partitions graphs into individual subproblems and assigns each subproblem to a suitable quantum solver based on their resource estimations. HyDRA-MIS is benchmarked against KaMIS and other hybrid quantum algorithms and performs competitively. We further demonstrate the scalability of our algorithm for future generations of quantum hardware.
\end{outline}

The paper is structured as follows. Section~\ref{sec:background} includes the technical background for GATs and solving MIS problems with QAOA. In Section~\ref{sec:methods} we describe the GAT-based parameter transfer scheme for MIS QAOA circuits, as well as HyDRA-MIS for solving large MIS instances. In Section~\ref{sec:results}, we analyze the performance of GAT-based QAOA parameter transfer for graphs of varying sizes and densities as well as benchmark HyDRA-MIS, and in Section~\ref{sec:discussion}
we conclude with future research directions.
\section{Background}\label{sec:background}
\subsection{Graph Attention Networks}
The Graph Attention Networks (GAT) are a special type of Graph Neural Networks (GNN) which can learn from graphs where feature representations for each node are aggregated over its neighborhood. In typical GNNs, representations for a node $i$ are updated through weighted averages over $i$'s neighborhood, where the weights depend only on the degree of $i$ which is similar to spectral embeddings \cite{luo2003spectral,chen2011algebraic}. In GATs, we update the feature representations of $i$ with Eq.~\ref{eq:gat}
\begin{equation}\label{eq:gat}
    h_i' = \sigma (\sum_{j\in N(i)} \alpha_{i,j} \mathbf{W} h_j),
\end{equation}
where $\sigma$ is any non-linear activation function, e.g., the exponential linear unit (ELU) function, and $\mathbf{W}$ is the learnable weight matrix. The attention weight $\alpha_{i,j}$ represents the importance of neighbor nodes' features to node $i$ during message passing and aggregation, and is defined for each edge $(i,j)$ in the graph according to Eq.~\ref{eq:attention_weight}
\begin{equation}\label{eq:attention_weight}
    \alpha_{i,j} = \text{softmax}_j(e_{i,j}),
\end{equation}
where $e_{i,j}$ denotes the attention score between node $i$ and $j$, which is also learned during training. The attention mechanism assigns different importance to different neighbors, thus allowing the network to concentrate on relevant parts of the graph. Multiple formulations for the attention score, $e_{i,j}$, exist in literature, including the original GAT paper~\cite{velickovicGraphAttentionNetworks2018} and GATv2~\cite{brodyHowAttentiveAre2022}. Interested readers are referred to both papers for details on implementations.

\subsection{Solving MIS with QAOA}
The QAOA is a hybrid algorithm with one of its main  applications related to solving combinatorial optimization problems. In its original form~\cite{farhiQuantumApproximateOptimization2014}, the circuit involves a mixing Hamiltonian $H_M= \sum_{i\leq n}\sigma_i^x$, and a problem Hamiltonian $H_P$, whose ground states encode the solutions to the original optimization problems. The QAOA circuit consists of $p$ layers of repeating parameterized ansatz:
\begin{equation}
    \ket{\psi_p(\beta,\gamma)} = e^{-i\beta_p H_M}e^{-i\gamma_p H_P}\dotsc e^{-i\beta_1 H_M}e^{-i\gamma_1 H_P} \ket{\psi_0},
\end{equation}
which is analogous to a Trotterized approximation to Quantum Adiabatic Algorithms~\cite{albashAdiabaticQuantumComputation2018}. The initial state $\psi_0$ is usually set to be the ground state of $H_M$ for the best performance~\cite{heAlignmentInitialState2023}, and one standard choice is $\ket{\psi_0}=\ket{+}^{\otimes n}$. The goal is to find a set of parameters $\beta_i$'s and $\gamma_i$'s for all $i$ such that the expected objective in Eq.~\ref{eq:exp_qaoa_obj} is minimized.
\begin{equation}\label{eq:exp_qaoa_obj}
    E(\gamma_1,\beta_1,\dotsc,\gamma_n,\beta_n) = \bra{\psi_p}H_P\ket{\psi_p} 
\end{equation}
Usually, the parameters are optimized using classical algorithms over several iterations, which can be time-consuming given the nonconvex and nonlinear nature of Eq.~\ref{eq:exp_qaoa_obj}. 

A typical class of problems that can be encoded in $H_P$ is the quadratic unconstrained binary optimization (QUBO) problems written in the following form:
\begin{equation}\label{eq:qubo_model}
    \min_{x} \ Q(x) = \sum_{i} h_i x_i + \sum_{i,j}J_{ij}x_ix_j,\quad  x_i\in\{0,1\}.
\end{equation}
The general QUBO minimization problem is NP-hard~\cite{barahonaComputationalComplexityIsing1982} and constrained problems can be converted to QUBOs with techniques such as Lagrangian relaxation. 

In this work, we solve the (unweighted) MIS problem using its standard QUBO formulation in Eq.~\ref{eq:mis_qubo}
\begin{equation}\label{eq:mis_qubo}
    Q(x_1,\dotsc,x_n) = -\sum_{i \in V} x_i + J \sum_{(i,j)\in E} x_i x_j.
\end{equation}
The variable $x_i = 1$ if and only if vertex $i$ is included in the solution. $-\sum_{i \in V} x_i$ counts the number of vertices in the independent set and $J\sum_{(i,j)\in E} x_i x_j$ penalizes for $J$ when both endpoints of an edge are included. The formulation is optimal for any $J > 1$~\cite{choiEffectsProblemHamiltonian2020}. 

\subsection{Related Work  on QAOA Parameter Transfer}
Identifying efficient methods for computing good QAOA parameters has gained significant interest due to the difficulty of direct optimization methods. Wurtz et al. compute a set of fixed parameters for MaxCut QAOA circuits, which are shown to be universally good in any regular graph of degree 3, 4 and 5 with constant edge weight~\cite{wurtzFixedAngleConjecture2021}. They numerically confirm that fixed parameters can usually obtain solutions with  approximation ratios close to those from optimal parameters. In \cite{streifTrainingQuantumApproximate2019}, instead of evaluating QAOA circuits on quantum computers during parameter optimizations, the authors introduce a contraction scheme for the tensor network representing arbitrary QAOA correlation functions, which can be computed efficiently on classical computers. Quantum computers are only queried after parameters are optimized, saving time and resources.  

Graph structures are also used for predicting parameter transferability between problem instances. For MaxCut problems, Galda et al. find success in transferring parameters between graphs with similar subgraph components, even as the graphs vary greatly in sizes~\cite{galda2023similarity,galdaTransferabilityOptimalQAOA2021}. For weighted MaxCut problems with significantly different objective landscapes, Shaydulin et al. \cite{shaydulinParameterTransferQuantum2023} propose a simple rescaling scheme to transfer parameters from unweighted MaxCut instances to weighted instances. The method works on edge weights generated from varying distributions and performs competitively with those found by direct methods.

Algorithms described above focus on parameter transfer for MaxCut problems and the Sherrington-Kirkpatrick (SK) model. However, practical problems are often constrained, which limits the subspace of feasible solutions and may break assumptions of existing techniques. Therefore, we propose a novel parameter transfer scheme for solving MIS problems which use GATs to learn the hidden connections between instances.

\subsection{Previous Quantum Solvers for Large MIS Instances}
Another obstacle for using QAOA to solve practical problems is the limitation in hardware size. Divide-and-conquer (DC) is a popular approach for solving large MIS instances on NISQ computers. Tomesh et al. propose a DC algorithm for solving combinatorial optimization problems on distributed quantum architectures~\cite{tomeshDivideConquerCombinatorial2023}. The algorithm partitions the graph into smaller subgraphs and selects a set of `hot' vertices from the boundary that remains entangled after circuit cutting. Computation time for post-processing scales exponentially with the number of hot vertices. Ibrahim et al. have later scaled this approach up in \cite{cameron2024scaling}. In \cite{xuDivideConquerbasedQuantum2024}, the authors use vertex separators to partition the graphs and solve the MIS problem on each subgraph separately. The separator-based DC algorithm is designed for separable graphs, which include planar graphs~\cite{liptonSeparatorTheoremPlanar1979}, finite element meshes~\cite{millerSeparatorsTwoThree1990} and certain classes of geometrically defined graphs~\cite{Miller+:JACM}. 

Our proposed algorithm, HyDRA-MIS, partitions the graph into smaller subproblems, similar to the approach of Tomesh et al., but uses simple classical post-processing to combine the solution. Instead of terminating after one pass as in \cite{xuDivideConquerbasedQuantum2024}, we compute different contiguous partitions for the graph and iteratively refine the solution. We compare HyDRA-MIS with KaMIS, as well as the separator-based DC algorithm in Section~\ref{sec:results}.

\section{Methods}\label{sec:methods}
\subsection{GAT for Transferring QAOA Parameters}\label{subsec:parameter_transfer}
Let us describe the procedures of using GATs to transfer QAOA parameters for MIS problems:
\subsubsection{Training Preparations}
The training graphs, or graphs which we will transfer parameters from, consist of Erd\"{o}s-R\'{e}nyi random graphs with 12 and 14 vertices and varying densities. We construct a two-layer MIS QAOA circuit for each graph and optimize their parameters from scratch. Multiple sets of parameters are computed for each graph, all starting with different initial solutions. We also train a \textit{graph2vec}~\cite{narayananGraph2vecLearningDistributed2017} model for the training graphs, which will be used later for parameters transfer. 

Furthermore, we compute a distance matrix for each pair of graphs, $(G_i, G_j)$, where the distance is defined as the `transferability' between their corresponding QAOA circuits. First, we pick the top five parameters from $G_i$'s circuit in terms of expected energy, transfer them to $G_j$'s circuit and record the best MIS approximation ratio. Second, we repeat the same process by transferring $G_j$'s parameters to $G_i$ and set the distance between $G_i$ and $G_j$ to be the average of these two ratios. 

\subsubsection{GAT Training}
We perform hierarchical clustering on the training graphs using the distance matrix computed from the previous step. Each graph is then assigned a corresponding cluster label for classification. For each graph, we compute their \textit{node2vec}~\cite{groverNode2vecScalableFeature2016} embeddings and use these as input features for GAT. Node features for each graph are aggregated and averaged after GAT convolution layers, resulting in a single vector as the full graph embedding. Lastly, we attach a standard multilayer perceptron in the end, which returns a probability distribution of potential cluster labels.

\subsubsection{Parameter Transfer}
Given a graph $H$, we predict its cluster label using the trained GAT model. Afterward, we find the closest $k$ training graphs in the cluster to $H$ in terms of their Euclidean distances in graph2vec embeddings. We transfer each of these $k$ graphs' best parameters to $H$, construct a MIS QAOA circuit and sample for 5,000 shots. The best feasible solution is then returned. In Section~\ref{sec:results}, we set $k=3$ for benchmarks. 

\subsection{Hybrid Distributed Resource-Aware Algorithm for MIS}
In order to solve large MIS instances that do not fit NISQ computers, we propose a hybrid distributed resource-aware algorithm, or HyDRA-MIS, and provide the pseudocode in Algorithm~\ref{alg:mis_quantum}. In the first iteration, we partition the graph into $k$ contiguous parts and use QAOA to solve MIS on each part in parallel. The combined solution is then repaired by greedily processing conflicting edges (i.e., those that violate MIS constraints). 

In the next iterations, we start with a different partition $P$ and attempt to improve the solution $S$ obtained previously: for every partition $Q_i$, we first remove vertices from $S$ if they are also present in $Q_i$. Then, we remove boundary vertices from $Q_i$ which have neighbors in $S$, solve the MIS problem on its induced subgraph and add the solution to $S$. We repeat this process for a fixed number of iterations or terminate early if no improvement can be made. In a distributed setting with heterogeneous quantum architecture, instead of processing partitions sequentially, we collect partitions with no cross-edges, and process them concurrently. For each subproblem, we estimate how much resource it will take and assign it to a suitable quantum solver. For example, dense subproblems will be assigned to ion-trapped quantum computers due to their high connectivity, while sparse subproblems will be assigned to superconducting quantum computers.

\begin{algorithm}[!htb]
\caption{Hybrid MIS Solver using QAOA}\label{alg:mis_quantum}
\begin{algorithmic}
\Function{Solve\_MIS}{$G = \{V,E\}$, $n$, $k$}
\State // \textit{$n$ dictates the total number of iterations and $k$ sets the number of partitions}
\State $S \gets$ MIS\_Initial\_Sol($G$, $k$)
\For {$n$ iterations}
    \State $S \gets $ MIS\_Iterative\_Sol($G$, $S$, $k$)
\EndFor
\State \Return $S$
\EndFunction
\end{algorithmic}
\begin{algorithmic}
\Function{MIS\_Initial\_Sol}{$G = \{V,E\}$, $k$}
\State Partition $V$ into $k$ balanced, contiguous parts
\State $S \gets$ combined MIS solution from solving each partition
\State $S \gets$ Repair\_MIS\_Sol($G$, $S$)
\State \Return $S$
\EndFunction
\end{algorithmic}
\begin{algorithmic}
\Function{Repair\_MIS\_Sol}{$G = \{V,E\}$, $S$}
\State $Q \gets$ set of edges which have both endpoints in $S$
\State $H=\{V_H, E_H\} \gets$ subgraph of $G$ induced by $Q$
\For {$v \in V_H$ in increasing order of degree}
    \If {$\exists u \in V_H \text{ such that } (u,v) \in E_H$}
        \State Remove $v$ from $S$
        \State Remove $v$ from $H$ and update $V_H$ and $E_H$
    \EndIf
\EndFor
\State \Return $S$
\EndFunction
\end{algorithmic}
\begin{algorithmic}
\Function{MIS\_Iterative\_Sol}{$G = \{V,E\}$, $S$, $k$}
\State Compute a different $k$-contiguous partition $P$ of $V$
\For {$Q_i$ in $P$}
    \For {$s\ in S$}
    \If {$s \in Q_i$}
        \State Remove $s$ from $S$
    \EndIf
    \EndFor
    \For {each boundary node $v \in Q_i$}
        \If {$\exists u \in S$ such that $u \notin Q_i$ and $(u,v) \in E$}
            \State Remove $v$ from $Q_i$
        \EndIf
    \EndFor
    \State $S' \gets$ solve MIS of the induced subgraph from $Q_i$
    \State $S = S + S'$
\EndFor
\State \Return $S$
\EndFunction
\end{algorithmic}
\end{algorithm}
\section{Computational Results}\label{sec:results}
\subsection{Experimental Setup}
We generate 3,000 random graphs of 12 and 14 vertices for training. Densities of the graphs are evenly distributed, ranging from $0.1$ to $0.9$. We use \textit{Pennylane}~\cite{bergholmPennyLaneAutomaticDifferentiation2022} to simulate and optimize parameters for QAOA circuits and generate more than 200 sets of parameters for each graph using different initial solutions obtained via grid search.

For HyDRA-MIS, we use \textit{METIS}~\cite{Karypis_1998} to partition graphs into contiguous parts. To guarantee partitions are different at each iteration, we assign random weights to edges of the graphs. The graph partitioning solver implemented in \textit{METIS} minimizes the total weight of cutting edges which ensures a randomization of partitions. We choose the number of partitions $k$ such that the largest partition does not exceed 25 vertices. For benchmarks, we use finite element meshes from \textit{SuiteSparse}~\cite{davisUniversityFloridaSparse2011} and compare MIS solution cardinalities against the separator-based DC algorithm~\cite{xuDivideConquerbasedQuantum2024}. We disable resource estimations in HyDRA-MIS, as state vector simulators are used in place of quantum computers.

\subsection{GAT-based QAOA Parameter Transfer Scaling}
\begin{figure}
    \centering
    \includegraphics[width=\linewidth]{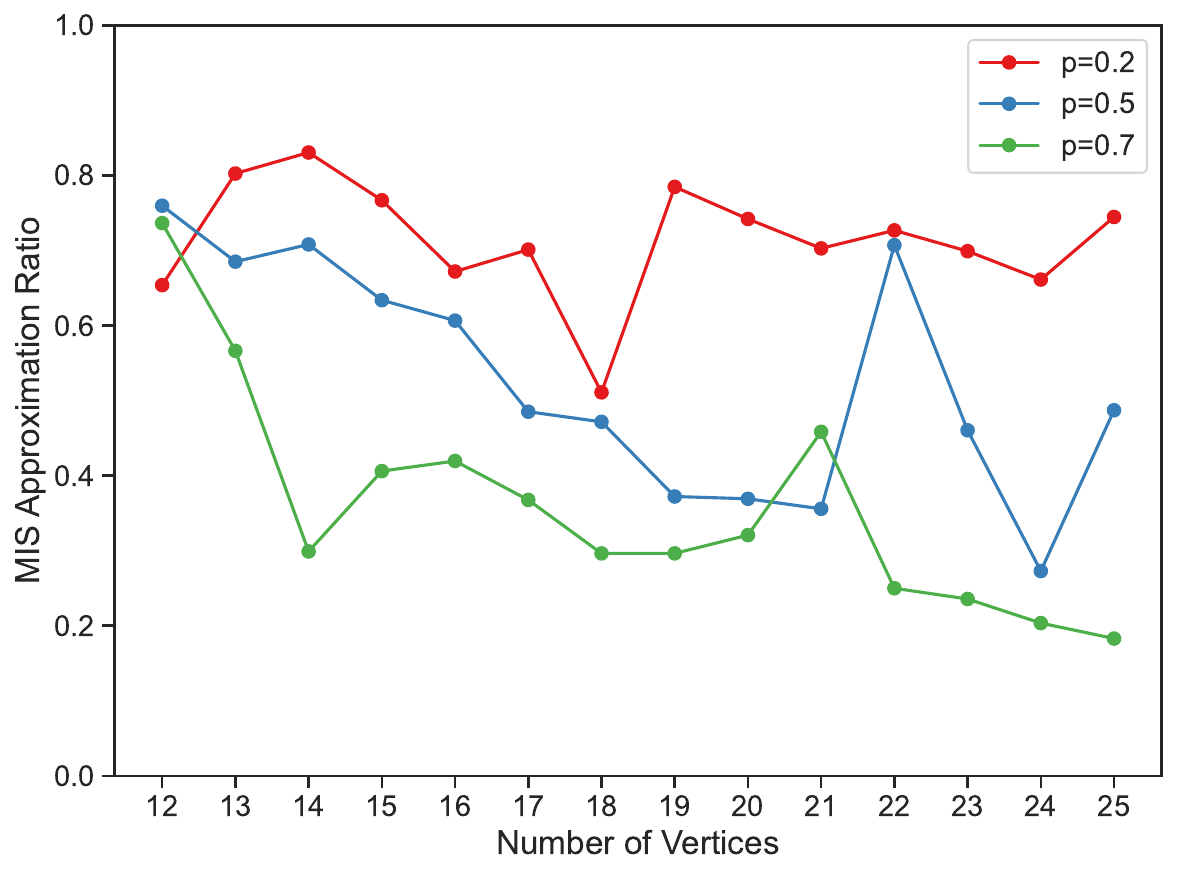}
    \caption{Approximation ratios of feasible samples from MIS QAOA circuits on graphs with varying sizes and densities. The parameters for each circuit are transferred via GATs.}
    \label{fig:graph_size_scaling}
\end{figure}

In Figure~\ref{fig:graph_size_scaling}, we generate 42 Erd\"{o}s-R\'{e}nyi random graphs with different number of vertices (12 to 25) and densities (0.2, 0.5 and 0.7). We apply our GAT-based parameter transfer scheme described in Section~\ref{subsec:parameter_transfer}, sample for 5000 times and record MIS approximation ratios over feasible samples. In general, we obtain better solutions on small sparse graphs, which also aligns with the findings in \cite{cazalsIdentifyingHardNative2025}. 

\subsection{Comparison of HyDRA-MIS}
\subsubsection{Scaling Comparison}
\begin{figure}
    \centering
    \includegraphics[width=\linewidth]{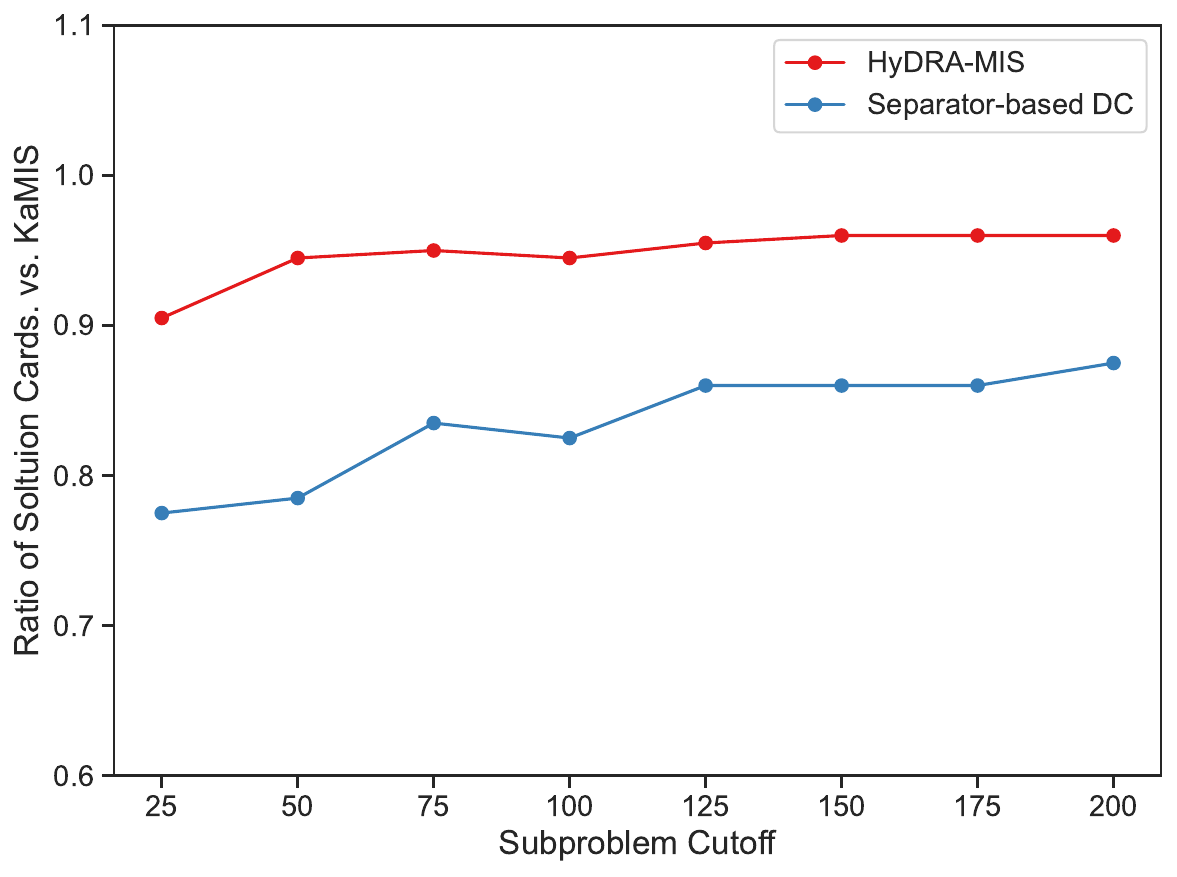}
    \caption{Comparison of independent set cardinalities obtained from HyDRA-MIS and separator-based DC algorithm with different subproblem cutoffs. Each subproblem is solved using KaMIS since simulating QAOA circuits with more than 50 qubits is beyond our capabilities.}
    \label{fig:algorithm_scaling}
\end{figure}

In Figure~\ref{fig:algorithm_scaling}, we compare MIS solutions obtained from running HyDRA-MIS and the separator-based DC algorithm on the graph \textit{bcsstk28} - one of the hardest instances in our test set. HyDRA-MIS is able to obtain an independent set of the size 95\% of that from KaMIS, even when the subproblem size is under 100 vertices. The DC algorithm, however, fails to find a solution that reaches 90\% of KaMIS'.

\subsubsection{Experimental Results}\label{subsec:distributed_result}
\begin{figure}
    \centering
    \includegraphics[width=\linewidth]{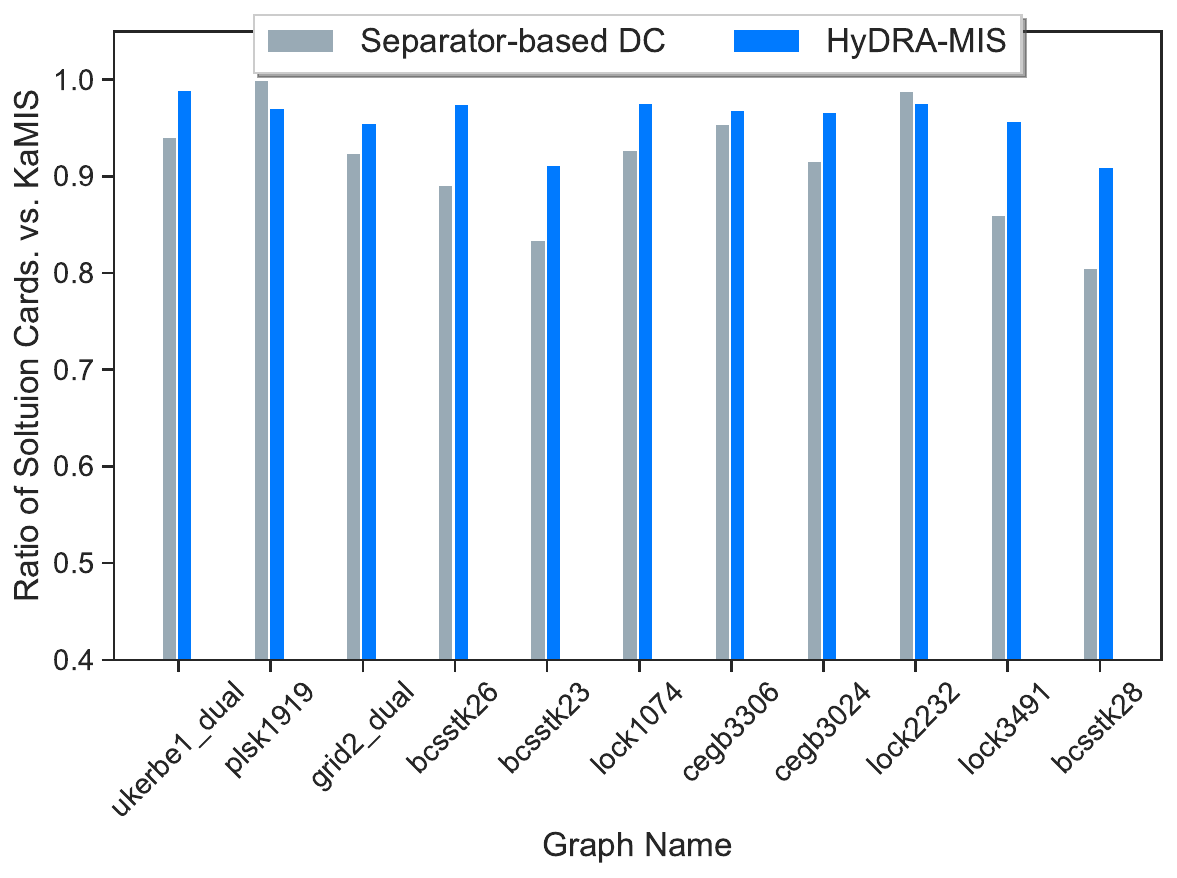}
    \caption{Comparison of independent set cardinalities obtained from HyDRA-MIS and separator-based DC algorithm. The y-axis tracks the ratios of solution cardinalities against the baseline solver, KaMIS. For each graph, we run KaMIS ten times and record the best solution.}
    \label{fig:mis_benchmark}
\end{figure}

Here we benchmark HyDRA-MIS against the DC algorithm on 11 graphs. Descriptions of the test graphs are provided in Table~\ref{tab:suitesparse}. The maximum subproblem size is set to 25 in both algorithms, and HyDRA-MIS terminates after 20 iterations. We normalize the results using solutions obtained from KaMIS, and report the comparison in Figure~\ref{fig:mis_benchmark}. HyDRA-MIS leads in 9 out of 11 instances, where \textit{lock3491} and \textit{bcsstk} graphs demonstrate the highest improvements. 

\begin{table*}[ht]
\caption{Statistics for graphs used in the experiments from \textit{SuiteSparse}. For each problem we report  its number of vertices, number of edges, maximum degree ($\Delta$), average degree, standard deviation in the degree and concurrency, which is defined as the maximum number of concurrent subproblems (disjoint partitions) during a 20-iteration run.}\label{tab:suitesparse}
\centering
\begin{tabular}{| c | c | c | c | c | c | c | c |}
\hline Name  & Planar & $\vert V\vert$ & $\vert E\vert$ & $\Delta$ & Avg Deg & Deg Std & Concurrency \\ 
\hline 
 ukerbe1\_dual & Y & 1866 & 3538 & 4 & 3.79 & 0.41 & 27\\
 \hline 
 plsk1919 & Y & 1919 & 4831 & 6 & 5.03 & 1.16  & 25\\
\hline 
 grid2\_dual & Y & 3136 & 6112 & 4 & 3.90 & 0.31 & 37 \\
 \hline 
 bcsstk26 & N & 1922 & 14207 & 32 & 14.78 & 5.69 & 26 \\
 \hline
 bcsstk23 & N & 3134 & 21022 & 30 & 13.42 & 4.87 & 24 \\
 \hline
 lock1074 & N & 1074 & 25275  & 95 & 47.07 & 15.88 & 13 \\
 \hline 
 cegb3306 & N & 3306 & 35847 & 53 & 21.67 & 11.25 & 47 \\
 \hline 
 cegb3024 & N & 3024 & 38426 & 67 & 25.41 & 9.37 & 31 \\
 \hline 
 lock2232 & N & 2232 & 39072 & 47 & 35.01 & 11.20 & 21\\
 \hline 
 lock3491 & N & 3491 & 78514 & 112 & 44.98 & 12.88 & 36 \\
 \hline 
 bcsstk28 & N & 4410 & 107307 & 93 & 48.67 & 9.68 & 42 \\
 \hline 
\end{tabular}
\end{table*}

\subsubsection{Future Outlook}
\begin{figure}
    \centering
    \includegraphics[width=\linewidth]{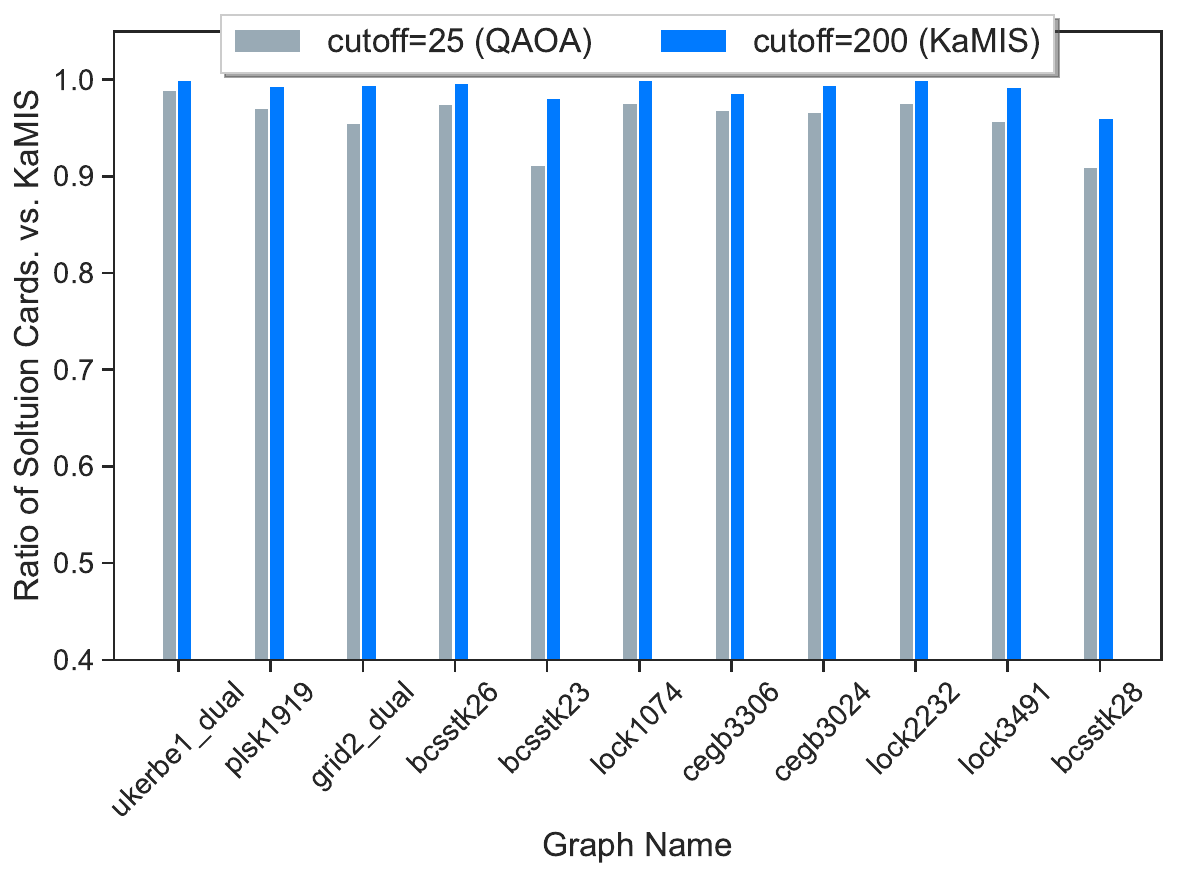}
    \caption{Comparison of independent set cardinalities obtained from HyDRA-MIS using different cutoff sizes. Subproblems are solved via QAOA simulation (cutoff=25) and KaMIS (cutoff=200).}
    \label{fig:future_outlook}
\end{figure}

To simulate the performance of HyDRA-MIS in future generations of quantum computers, we perform the same benchmarks as in Section~\ref{subsec:distributed_result} but set the cutoff to 200 vertices. Each subproblem is solved via KaMIS as the circuits are out of reach for classical state-vector simulations. We compare the results to those from QAOA simulations with a cutoff of 25 in Figure~\ref{fig:future_outlook}. Using a cutoff size of 200, we can obtain independent sets with sizes over 99\% of those from KaMIS in 9 out of the 11 instances, providing evidence for scalability using future quantum hardware.

\section{Discussion}\label{sec:discussion}
In this paper, we propose a QAOA parameter transfer scheme for Maximum Independent Set problems using Graph Attention Networks. In order to better utilize such techniques in NISQ era where quantum computers are constrained by sizes, noise and coherence times, we design HyDRA-MIS, a hybrid distributed resource-aware algorithm which partitions the original graph into many disjoint subgraphs. We benchmark HyDRA-MIS on sparse finite elements meshes up to 4,410 vertices, enabling distributed computing on subgraphs. Furthermore, we report results that are comparable to KaMIS, a state-of-the-art classical MIS solver, and demonstrate clear advantages over the previous separator-based DC algorithm in the majority of the instances.

It is not surprising that distributed hybrid quantum-classical algorithm for MIS generally do not reach the same solution quality as its sequential counterpart KaMIS. This discrepancy arises from the inherent constraints of distribution: the global problem must be partitioned into smaller subproblems that are solved independently, often with limited coordination and communication among computational units. As a result, critical global dependencies and interactions between distant parts of the graph may be lost or approximated during partitioning, leading to suboptimal boundary handling and reduced solution coherence. While distributed approaches offer scalability and are better suited for large-scale or hardware-limited environments, they inherently trade off some solution quality for parallel efficiency and feasibility on near-term quantum architectures.

In the experiments, we use predominantly finite element meshes in order to compare with the separator-based DC algorithm. Sparse graphs also enable concurrent subproblem processing for HyDRA-MIS, as shown in Table~\ref{tab:suitesparse}. For dense graphs, however, it may be harder to find partitions with no cross-edges. Thus, one direction for future research is to scale the algorithm for a more general class of graphs. Other directions include exploring alternative GAT constructions with different node feature representations and additional edge features. More training samples over a wider range of graphs may also be necessary for improving parameter transfer accuracy. On the side of QAOA, it is worth investigating parameter transfer for QAOA circuits using constrained mixers~\cite{fuchsConstrainedMixersQuantum2022}, which restrict the state evolution to a feasible subspace. It is also interesting to apply both the parameter transfer scheme and distributed framework to other graph problems, including Maximum k-Colorable Subgraph and Minimum Dominating Set problems.

\section*{Acknowledgment}
H.X. is supported by the Center for Quantum Technologies (CQT) at Purdue University, which is an Industry-University Cooperative Research Center (IUCRC) funded through the US National Science Foundation (NSF) under Grant No. 2224960. 
I.S. was partially supported with funding from the Defense Advanced Research Projects Agency (DARPA)  ONISQ program and by National Science Foundation ExpandQISE program award \#2427042. 
We also thank Siddhartha Das for his valuable insight into GATs.
\bibliographystyle{IEEEtran}
\bibliography{bib,ilya-biblio}

\end{document}